\documentclass[12pt]{article}
\input xy
\xyoption{all}
\input epsf.tex
\usepackage{pstricks}
\usepackage{psfrag}
        
\usepackage{graphicx}
\usepackage[small,loose]{subfigure}
\usepackage{mathrsfs}
\usepackage{bbm} 
\usepackage{cancel}
\usepackage{amssymb,amsmath}
\def\harr#1#2{\smash{\mathop{\hbox to .3in{\rightarrowfill}}
 \limits^{\scriptstyle#1}_{\scriptstyle#2}}}

\def\s2{\frac{1}{\sqrt2}}

\def\be{\begin{equation}}
\def\ee{\end{equation}}
\def\beqa{\begin{eqnarray}}
\def\eeqa{\end{eqnarray}}

\def\Dsl{\,\raise.15ex\hbox{/}\mkern-13.5mu D} 
\def\d3{d^3}

\begin{document}

\vspace{.5cm}
\begin{center}
\Large{\bf  Dynamic Hurst Exponent in Time Series   }

\vspace{1cm}

\large Carlos Arturo
Soto Campos\footnote{e-mail address: {\tt
csoto@uaeh.edu.mx}} \\
[2mm] 
{\small \em Universidad Aut\'onoma del Estado de Hidalgo}\\
{\small\em Carretera Pachuca-Tulancingo  Km.4.5\\
Pachuca, Hidalgo}
\\[4mm]

\large Leopoldo S\'anchez Cant\'u\footnote{e-mail address: {\tt
polo.antares@gmail.com}}\\
[2mm] 
{\small \em Center of Complexity Sciences, Unam.}\\
{\small\em Cd. Universitaria, Delegaci\'on Coyoac\'an, 04510\\
Ciudad de M\'exico}\\[4mm]

\large Zeus Hern\'andez Veleros\footnote{e-mail address: {\tt
zshveleros@yahoo.com}}\\
[2mm] 
{\small \em Universidad Aut\'onoma del Estado de Hidalgo}\\
{\small\em San Agustín Tlaxiaca Hidalgo, M\'exico
C.P. 42160\\
Pachuca, Hidalgo}
\\[4mm]

\vspace*{1.5cm}
\small{\bf Abstract} \\
\end{center}

\begin{center}
\begin{minipage}[h]{14.0cm}
{The market efficiency hypothesis has been proposed to explain the behavior of time series
of stock markets. The Black-Scholes model (B-S) for example, is based on the assumption that markets are efficient. As a consequence, it is impossible, at least in principle, to ``predict" how a market behaves, whatever the circumstances. 
Recently we have found evidence which shows that it is possible to find self-organized 
behavior in the prices of assets in financial markets during deep falls of those prices. 
Through a kurtosis analysis we have identified a critical point 
that separates time series from stock markets in
two different regimes: the mesokurtic segment compatible with a random walk regime and the leptokurtic one 
that allegedly follows a power law behavior. In this paper we provide some 
evidence, showing that the Hurst exponent is a good estimator of the regime in which the market is operating. Finally, we
propose that the Hurst exponent can be 
considered as a critical variable in just the same way as magnetization, 
for example, can be used to distinguish the phase of a magnetic system in physics.    
}
\end{minipage}
\end{center}

\bigskip
\bigskip

\date{\today}

\vspace{3cm}


\newpage

\section{Introduction}

Phase transitions are one of the most subtle ideas in physics. Despite its apparently simple explanation, the different behavior of water in the vicinity  of the triple point still represents a challenge from the perspective of an average physics student. Leon Kadanoff has pointed out that, ``A phase transition is a change from one behavior to another. A first order phase transition involves a discontinuous jump in some statistical variable. The discontinuous property is called the order parameter. Each phase transition has its own order parameter." \cite{Kadanoff1}

Now, much work has been done  in the search of a realistic theory that could
explain the behavior of real-life stock market. In this line of thinking, the work of Kadanoff seems particularly adequate because it exposes the relationship between physical and economic phenomena in terms of the so-called Mean Field Theory (MFT). Citing the work of Kadanoff already mentioned before: 

``Economic theory, as first worked out by Adam Smith and then further expounded by Paul Samuelson, has an equilibrium model that treats money in much the same manner as Maxwell and Boltzmann treated energy in their development of statistical mechanics... From this theory, one could deduce a model predicting a \textit{rational} price for a derivative security. This Black-Scholes (B-S) model had some of the same content as the mean field theory models used to describe phase transitions." \cite{Kadanoff1}

So, the proposed B-S financial model attempted to calculate the future value of a derivative security by taking averages assuming that the statistical distribution is ``well behaved", which means, a gaussian \cite{black-scholes}. As in the case of MFT, here the fluctuations were treated approximately, and correlations among fluctuations of different variables were ignored. 

It is not surprising then that the B-S model had failed to reproduce the behavior of assets prices in a real stock market, especially when correlations between variables were not properly included in the original model.

\vskip 0.8truecm
\section{Some comments on Critical Phenomena}

A very well known system that presents a transition at a critical point is a ferromagnet. Ferromagnets have a characteristic behavior because they have a permanent magnetization $M$. When the temperature is raised to the critical point $T_c$ (also known as the Curie point) the magnetization is proportional to the magnetic field, the system ceases to be ferromagnetic and becomes paramagnetic
$$ \frac{M}{H} = \frac{C_c}{T - T_c}   $$
Where $C_c$ is a constant depending on the atomic properties of each particular substance. The form of this equation shows that at the Curie point we expect a singularity in the ratio $M/H $. Since the maximum value of $M$ is finite we must conclude that $H = 0$. That means that the system is magnetized even in the absence of an external magnetic field. That is the characteristic of a ferromagnet. 
The Curie temperature $T_c$ represents the boundary between paramagnetic and ferromagnetic behavior depending if the temperature $T$ is bigger or smaller than the critical value. This situation is analogous to the case of a fluid. The increment of the pressure $p$ of a fluid increases the density $\rho$ and the increment of the magnetic field $H$ to a magnet increases the magnetization $M$. Thus, the pressure $p$ and the magnetic field $H$ play a similar role. The same happens with the density $\rho$ and the magnetization $M$, both variables play a similar role in different systems.  

\medskip

Now, let us review the approach of Statistical Mechanics, a theory that attempts to describe the behavior of many bodies interacting with each other. In that theory, the so-called critical exponents have revealed the structure of fluctuations at many scales nesting with one another. Microscopic parameters affect the structure of fluctuations only at the microscopic scale, and those fluctuations are quickly averaged at the macroscopic level. Consequently, microstructures do not affect the various ways in which many systems approach their respective critical temperatures, this can be described by a few universal critical exponents and simple scaling laws. 

Nowadays, it is a common knowledge that water undergoes phase transitions as we vary some parameters as pressure, temperature and volume. For other substances, the different parameters used to describe the system could include magnetic fields, for instance. It does not matter what system we are trying to describe, its thermodynamics could be characterized by an analytical function, called the free energy per unit volume, $G(T, P, H, ...)$ or Gibbs free energy, which depends on the control parameters. It was the physicist Paul Ehrenfest who first defined the order of a transition as the order of the derivative that becomes singular or discontinuous at the phase transition. However, the understanding of the way that renormalization arise came first from the work of Kadanoff, Fisher and Wilson \cite{Kadanoff2, Fisher, Wilson1, Wilson2}. They were working on the problem of phase transitions and critical phenomena. 

Lev D. Landau \cite{Landau} realized that the main difference occurred between first-order phase transitions and all the other orders. He stated that the different phases of a system were characterized by one parameter $\Phi$ called the order parameter. In the case of a liquid-gas transition, this parameter would be related to the difference between the density of the water at some particular phase and the value of the critical density at which the phase transition occurs. 

We define the order parameter so that its value will be small near the transition. It must not have thermal fluctuations and it will be determined by minimizing the free-energy $G(T, \Phi, ...)$ with respect to  $\Phi$ at some particular value of the control parameters. Now, near the critical point we can expand the free energy function as

$$G = a(T) \Phi^2 + b(T)\Phi^3 + c(T) \Phi^4 + \dots $$ 
where each coefficient will depend on the control parameters as well as on the temperature. 
If we take the first derivative of $G$, keeping all the terms and assuming that $c>0$, we can see how phase transitions work: $G$ is continuous but not its first derivative. Then, varying $T$ we can jump from one minimum to the other. 

One could explain second-order phase transitions by putting the coefficient $b\to0$ by fine-tuning some other parameter, or assuming that the system is symmetric under the transformation $\Phi \to -\Phi $. In that case we can see that, if $a$ vanishes at critical temperature $a(T_c) = 0$, then the two minima of the function $G$ would collapse into one point. Free energy would have square root branch points at $T-T_c$. 

Mean Field Theory, as Landau's theory was called, has the property that it depends only on the properties of the polynomials and not on the nature of the order parameters. This is also known as universality. In the case of continuous phase transitions that property is very important. Let us take for instance the case of the liquid-vapor phase diagram in the $P,T$ plane: it has the same critical exponent than the so-called Ising ferromagnet. Generally speaking, the exponents are different for different systems. This contrasting behavior is precisely what makes so interesting Landau's theory.  

On the other hand, in Quantum Mechanics we can define the correlation function as 
 
 \begin{equation}
 C(t) = \left<\widehat O(t) \ \widehat O(0)\right>
\end{equation}
 where $\widehat O(t)$ is some operator depending on time in the so-called Heisenberg picture. Then, it is possible to find the correlation function of any operator at different times. The interpretation of such a function is obvious. Besides, it helps to better understand the present approach. Take for instance the one-dimensional quantum harmonic oscillator. For that system, the ground state function is given by the wave function 
 
 $$
 \left<\widehat X \big| 0 \right> = \frac{1}{\pi^{1/4} x_0^{1/2}}\exp{ \Big[ -1/2 (x/x_0)^2  \Big] }
 $$
 where $x_0 = \sqrt{\hbar / m\omega}$ and $\omega$ is the frequency associated to the harmonic oscillator and $\hbar$ is Planck's constant. Let us calculate the correlation of the position operator $\widehat X $ at the initial time $t = 0$ and at a later time $t$, we find that
 
 \begin{eqnarray}
 C(t) & = & \left<0 \big|\widehat X(t)\widehat X(0) \big|0\right>  \nonumber\\
 {} & = & \frac{\hbar}{2m\omega} \cos\omega t
 \end{eqnarray}
and we can verify that correlation oscillates around its maximum value $\frac{\hbar}{2m\omega}$ as time goes on, that is, for $\omega t = 2 n \pi$ with $n\in \mathbb N$. Taking into account just half a period, correlation is maximum at $t= 0$ and after that, data becomes less correlated, until it reaches the value $t = \pi/ \omega$. This last value represents totally uncorrelated data.


\medskip

\medskip

There are another set of phenomena, observed by scattering light and neutrons of substances which undergo second order phase transitions. In this phenomena, the correlation function behaves in different ways at two different regimes. This could be explained assuming that, at large distances and near the critical point, the correlation functions of local quantities like the magnetization $M(x)$, obey scaling laws like 

 $$\left<M(x)M(y)\right> \approx {\left| x-y \right|}^{-2\beta} $$             
where $\beta$ is the critical exponent. By contrast,  away from the critical point, condensed matter systems generally have the so-called correlation length 

$$
\left<M(x)M(y)\right> \approx \frac{e^{-\frac{\left| x-y \right|}{l_c}} }{\left| x-y \right| }
$$   
where $l_c$ has the dimension of length. These phenomena, have been explained in terms of the Renormalization Group in Physics. However, it is not the purpose of the present work, to explain things in terms of that theory. The great perspicacity of L. Kadanoff, was to realize that, if critical phenomena had something to do with an infinite correlation length, then the microscopic dynamics of the system must be irrelevant, and there should be some kind of long wavelength effective theory, retaining that characteristic. 
 
There still remain some questions about a possible extension of Mean field theory to Economics or Financial markets. In the case of deep falls of asset prices in a stock market, we still do not have the possibility to predict the change of regime or to detect a trigger (if it happens to be one). It would be necessary to state the order parameter whose change could determine the phase transition. 

On the other hand, there is not an agreement on how to define appropriate free energy for time series representing a financial market. In the case of thermodynamic systems, this is quite simple. Normally, the Gibbs function $G = H - TS$ where $H$ is the enthalpy, $T$ the temperature and $S$ the entropy of a system, does work well.  
 
 \medskip

 In the case of Financial markets, we have not succeeded in creating an equivalent Mean Field Theory, capable of predicting the arising of critical points and phase transitions, despite the suggesting similarities between the dynamics of stock markets with some physical systems. In the next section, we provide a possible explanation in terms of the presence (or absence) of self-organization of the values of assets in time series.  
 
 \vspace{0.5cm}

This paper is organized as follows: In section 3 we establish the presence of self-organization on some time series representing a financial market.  Section 4 is devoted to the study of the so-called Hurst exponent. In section 5 we have summarized the procedure we followed and the results we obtained. 

Finally, in section 6 we give our concluding remarks.


\vspace{0.5cm}
\section{Presence of self organization on Time series}

\medskip

In Economics,  the conventional model represents the fluctuations of asset prices 
in financial markets as a random walk where each state is independent of previous ones. This 
obviously suggests that it would be impossible to distinguish different behaviors in such a time series.

However, recently, we have developed a procedure to study each one of the
price drops, starting from a given maximum level \cite{Sanchez}. We have explored a range in the state space in which the falls have been 
explained as a process that follows a Power Law,  finding evidence of self-organization in time series of stock market indices. The standard model of the market establishes that there are ``anomalies'' in the time series, meaning that when certain behaviors are found in time series that can not be explained by the standard model, those are said to be ``anomalies". In the present approach, it has not been considered {\it{a priori}} that there could be anomalies in order to include all patterns observed as legitimate behavioral conducts that a model should explain.


\medskip

Considering all the aforementioned, we developed a methodology Ref. \cite{Sanchez} in order to identify, in a series of daily closing prices each downward movement to a bottom point from a recent peak, followed by a rebound back to the previous ceiling, or back to the highest value registered in the previous six months, whichever was reached first. These cumulative negative returns or downfalls then became units of study. Using the series of negative returns as observables, we explored the possibility of identifying a range within the space of states in which a variable that corresponded to the lowest value accumulated during each downfall could be explained as a process that follows a power law. 

In the case of decreased prices in the stock market, while the system is still under the random regime, small changes allegedly generated by the response of some investors to exogenous information, perceived as negative, are easily absorbed and reverted by the contingent of optimistic participants who may be analyzing the same information but in a longer (or shorter) time frame. Perhaps they concluded that the recent descent in prices creates a favorable condition to increase their positions. However, if the downhill movement of prices continues, the sequence of successive perturbations increases the tension generated in the system following new small downward impulses. At a certain point, the pressure upon the contingent of buyers eventually overrides their capacity to absorb the increasing bidding until a further decrease in prices generates a change of regime (bifurcation) in which rather than buyers, new sellers who wish to get rid of their positions to stop their growing losses, are attracted. Thus, a positive feedback loop is built:  lowered prices attract more sellers whose bidding presses prices downwards in a vicious circle that generates a selling crisis characteristic of a market sell-off.

Data sets of downfalls of 30 international equity market indices (7 regional international, 5 American, 4 Latin American, 4 from emerging European countries, 5 from developed European countries and 5 from Asian countries) were ordered by size. The absolute value of each downfall (ordinates) was plotted against the place it occupied by size (abscissas) on a log-log scale.
Kurtosis of the progressive sets of absolute downfalls of each index was calculated, anchoring each series at the smallest decrease in value. Increasingly larger absolute decreased values were incorporated one by one until the largest was reached. The level of the one decrease from which the set of downfalls smaller to it had a kurtosis closest to zero was identified as the cutoff point or critical level. The set of events with kurtosis closest to zero is thus compatible with a normal density distribution.
The cutoff point was recognized as the critical level of phase transition separating two regimes of operation. It was assumed that the lower (mesokurtic) segment of price decreases operated under a random regime,  while the upper (leptokurtotic) segment of downfalls larger than the critical level can be explained as following a power law. This latter set of larger losses is incompatible with a normal density distribution. We propose this set of larger declines in value to be considered generated by a process under a self-organized regime. However, we did not provide a method in terms of which we could see how those different regimes arise. In the next section we continue with this approach, exploring the behavior of the so-called Hurst exponent. 
\section{ Some comments on Hurst Exponent}

In 1951, H. E. Hurst (1880-1978) published a method that detected non-periodic cyclic patterns in the floods of the Nile River, in which long periods of relative drought were followed by long periods of repeated flooding. It was employed as a measure of a long-range persistence of time series related to autocorrelations of different variables \cite{Hurst}. The value of the persistence, named the Hurst exponent (H exponent), is connected to the so-called fractal dimension, introduced by B. Mandelbrot in 1967 Ref{\cite{Mandelbrot}}. 

Mandelbrot and Wallis found this long term memory effect in various time series of capital markets and future markets. They named it as the Joseph phenomenon or Hurst effect (Hurst, 1951, 1965, Mandelbrot, 1968, 2004).

The Hurst exponent has been used in many research fields. From a long time ago, it has been used in finance. Some of the most well-known works are due to Edgar E. Peters \cite{Peters}. The H exponent takes values between 0 and 1,  and depending on its value, the time series can be classified into three different categories: if $H \in (0, 0.5)$ that implies that the series is anti-persistent. If $H =0.5$ this implies that the series behaves randomly. Finally, if $H \in (0.5, 1.0)$ this means that the series is persistent. This behavior becomes more persistent as the value of  $H$ approaches $1.0$. 

\medskip

More recently, in the paper Ref{\cite{Matos}}, Matos et al introduced the H exponent as a function of time and scale. Those authors claim that their method allows them to recover major effects affecting worldwide markets, and, at the same time, to analyze the way those effects propagate depending on different scales. They classify markets into mature and emergent ones. Some entities like the International Finance Corporation (IFC) uses national income per capita for classifying equity markets. It is a reference measure, giving information about a low or middle-income economy. If the ratio of the investable market capitalization to GNP is low, then IFC classifies the market as emerging. 

Matos et al begin by introducing fractional Brownian motion as the process in which 

$$
E \{{( X(t_2) -X(t_1)  )}^2  \}   \propto {|t_2- t_1  |}^{2H}
$$
where Brownian motion corresponds to $H = 1/2$. As usual, the authors stated that the values of $H$ could be greater or lower than $1/2$, depending on the {\textit{level of correlation }} of the values of the series.

\smallskip

\noindent Detrended Fluctuation Analysis (DFA) consists in dividing a
random variable sequence $X(t)$, of length $s$, into $s/ \tau$ non-overlapping boxes,
each containing $\tau$ points (5). The linear local trend $z(t) = a t + b$ in each box
is defined to be the standard linear least-square fit of the data points in that
box. The detrended fluctuation function F is then defined by: 

$$
F_k^2 = \frac{1}{\tau}\sum_{t=k+1}^{(k+1)\tau} {|X(t) - z(t) |}^2
$$
where $k = 0, ..., \frac{s}{\tau} -1 $. Averaging $F(\tau)$ over $s/\tau$ intervals, gives the fluctuation $\left<F(\tau) \right>$ as a function of $\tau$

$$
\left<F_k^2 \right> = \frac{\tau}{s} \sum_{\tau= 0}^{s/\tau-1} F_k^2(t)
$$
If the observables $X(t)$ are randomly uncorrelated variables or short-range correlated variables, then the behavior is expected to be 
that of a power law

$$
\left<F^2(\tau) \right> \approx \tau^H
$$
The authors studied the behavior of some particular examples, indices like the Nikkei of Japan, Bovespa of Brazil and the PSI20 of Portugal. In Ref(\cite{Matos}) the following classification is introduced: 

(i)(clearly) mature, this market have $H$ around 0.5. The presence of regions
with higher values of $H$ is limited to small periods and is well defined both
in time and scale.

\smallskip 

(ii)(clearly) emergent, this market have $H$ well above 0.5. The presence of
regions with values of $H$ around 0.5 is well defined both in time and scale.

\smallskip

(iii) hybrid, unlike the two previous cases, the distinction between the mature and
emergent phases are not well determined, with the behavior seemingly mixing at all scales.

\smallskip

Establishing in this way, different regimes, depending on the level of correlation of the variables.

\medskip

As we have emphasized, the $H$ exponent has recently acquired more relevance when we try to find long term correlations between different variables in time series representing the behavior of some financial market. In the same line as us, Carbone et al  Ref{\cite{Carbone}} calculated the $H$ exponent as a function of time, $H(t)$ of various time series by a dynamical implementation of the detrended moving average (DMA) in just the same way we have explained.   

\section{  Method and Results}

   \bigskip 


In the present work, the measurement of the Hurst exponent has been made using two methods. The first method was the measurement of the $H$ exponent of complete series of log returns, of absolute log-returns and of squared log-returns of 45 international stock market indices. The second one was the calculation of the aforementioned $H$ exponent by the standardized range analysis method of successive segments of the registered log-returns with windows of three sizes: 1000 days, 500 days and 250 days (window size = $\tau$). In all three window sizes, the study has been done on the complete series with 20 days lag (lag =  $\lambda$), making sweeps from the beginning of the data (many of them starting from the moment indexes were created) to the present time.

\smallskip

Next, we will describe the method we used and its justification.

\smallskip

We want to measure the strength of a trend that is embedded in a white noise environment.
For this, it has been necessary to determine the scaling exponent of the range of price movements as the observation scale is changed.

\smallskip

Mandelbrot proposed that, given the long-term dependency, perfect arbitrage would not be possible. He showed that the Brownian fractal movement was not a good model of stock returns unless the market is very inefficient, which is expanded by Hodges. The latter author calculates that for markets with an $H$ exponent outside the 0.4 to 0.6 range, fewer than 300 transactions would be necessary to obtain essentially risk-free profits. It provides a very useful table in which it relates the values of the Hurst's exponent with the Sharpe Index and the number of operations necessary to capture profits with options strategies \cite{Mandelbrot1971}. 

\smallskip

This Range Analysis Method at Different Scales ($R/S$ Analysis) allows us to calculate the value of the $H$ exponent according to the original procedure of Hurst, a method that we follow and describe here. 

\smallskip

We start with the log-returns analysis, $r_{t} = \ln p_{t} - \ln p_{t-1}$ where $r_{t}$ is the performance of a day in the day 
$t$, $p_{t}$ is the price in time $t$ and $p_{t-1}$ is the price one day before the day $t$. We will call $v$ each of the returns.

\smallskip

Let's assume a time series of size $N$, which we divide into $V$ length intervals $n$ such that $V_{n} = N$. Each interval of length $n$ is called $I_{\nu}$ in such a way that we have $\nu = 1, 2, ..., V$. Each element of the interval is called $N_{k, \nu}$  with $k = 1, 2, ..., n$.

\smallskip

\begin{enumerate}
	\item The average $m_{v}$ of the elements of each sub-interval of length $n$ is calculated obtaining $\nu$ measures calculated with the formula:
	
	\smallskip
	
	$$
	m_{\nu} = \frac{1}{n} \sum\limits_{k=1}^n N_{k,\nu}
	$$

	\smallskip

	\item The sample standard deviation is calculated for each subinterval $I_{v}$.
	
	$$
	S_{i,\nu} = \left(\sum\limits_{k=1}^n (N_{k,\nu}-m_{\nu})^{2}\right)^{\frac{1}{2}}
	$$

	\smallskip

	\item Cumulative deviations are calculated with respect to the mean for each subinterval.
	
	$$
	X_{k,\nu} = \sum\limits_{i=1}^n (N_{i,\nu}-m_{\nu})^{2}, \ \ k = 1,2,...,n
	$$

	\smallskip

	\item The range for each sub-interval $R_{Iv}$ is defined as the difference between the maximum and minimum value of $X_{k,v}$:

	$$
	R_{I\nu} = max\left(X_{k,\nu}\right) - min\left(X_{k,\nu}\right)  	$$

	\smallskip

	\item The ratio of this range divided by the standard deviation is obtained for each interval and the average ratio is calculated for each length interval set.
	$$
	(R/S)_{n} = \frac{1}{V} \sum\limits_{\nu=1}^V \left(R_{I\nu}/S_{I\nu}\right)
	$$
	
	\smallskip

	\item The length of the interval is increased to the next value that verifies that $N / n$ is always an integer and the process is repeated for all possible values of $n$.

	\smallskip

	\item A power regression is performed considering $\log(n)$ as the independent variable and $\log(R/S)_{n}$ as the dependent one. The exponent of the regression is the value of $H$.	
	\smallskip

	\item The calculations are made for the daily log-returns $r_{t} = \ln p_{t} - \ln p_{t-1}$, for the absolute of value of  the log-returns $\lvert r_{t} \rvert\ = \lvert \ln p_{t} - \ln p_{t-1} \rvert$ , and for the square of the returns $r_{t}^{2} = (\ln p_{t} - \ln p_{t-1})^{2}$ keeping the $R^{2}$ coefficient of each regression.
	
	\smallskip
	
	\item The autocorrelation function (the correlation between present and future values) can be calculated with the formula $C = 2^{2H-1} - 1$
\end{enumerate}

\smallskip

According to statistical mechanics, if the series is a random walk (without correlation and without memory), the $H$ exponent must be 0.5. In other words, if the process is purely random, the range of accumulated deviations must increase with the square root of time (if the exponent is $0.5$ then the scaling is given by an exponent of $1/2$, that is, it grows with $\sqrt{t}$, which in this case is $n$, that is, the number of successive data (log-returns, log-absolute returns or squared log-returns) that exist in each sample (each segment $v$, according to the previous notation). In addition, the correlation will be equal to $0$. 

\smallskip



	
	


What the exponent $H \neq 0.5$ detects, cannot be a short-term  (Markovian) memory but rather a long-term memory that tends to infinite. In other words, a system that exhibits Hurst-type statistics is the result of a long chain of interconnected events in which today's events influence events from the future. There exist a dependence on the path that has been followed.

\smallskip

Persistent time series (defined as those for which $0.5 <H <1.0$) are fractals since they can be described as a fractional Brownian motion. In the fractional Brownian movement there is a correlation between the events of different time scales. Hurst's exponent describes the probability that two consecutive events may recur. If $H = 0.6$, in essence, there will be a $60\%$ chance that if the last movement was positive, the next is also positive.

\smallskip

Since each event does not have the same probability of occurring (as in a random march), the fractal dimension of the probability distribution is not 2; but it is a number between 1 and 2. Mandelbrot showed that the fractal dimension $F_D$ is the inverse of  $H$ exponent, i.e. $F_D = 1 / H$.

\smallskip

The reduction of the phase space in the price path should be noticeable when having a fractal dimension less than 2, as it would be in the case of a random march.

\smallskip

It can be shown that the sum of the successive log-returns is equal to the composite return. In this procedure we are doing the same thing, but cutting into smaller and smaller pieces the object of study (the time series) and doing the sum around the mean.

\smallskip

Finally, we measure the size of the oscillations in each different scale (relative to the average of each segment in each scale) and we divide it by the volatility of that scale particular to ``standardize" the value of the oscillation, in such a way that the measurements of different scales are comparable to each other.

If the returns behave randomly, the scaling exponent should be equal $\sqrt t$, that means, $H =0.5$. When the value of $H$ is not 1/2, this implies that there is a non-random factor that determines the way of escalation. That factor is long-term memory of volatility, not of the returns themselves. Additionally, an exponent different from 0.5, evidences the fractal structure of the phenomenon. In conditions of randomness, the fractal dimension would be 2. Under non-random conditions, it would be between 1 and 2 ($1 <F_D <2$).

One of the most relevant properties of the $R / S$ analysis is that it represents a robust test for the detection of cycles. The search for cycles has been a constant in the economic study. By analyzing the quotient $R / S$ it is possible to detect non-periodic cycles whose period is greater than that of the sample and to know the approximate length of those cycles (Mandelbrot 1969). To do this, the use of the $V$ statistic, which is defined as follows:

$$
V_n = \frac{(R / S)_{n}}{\sqrt{n}}
$$

Keeping $\log (n)$ as the independent variable and $V_n$ as the dependent one, we will obtain the following result:

\begin{itemize}
	\item If the process is an ordinary Brownian motion, i.e. $H = 1/2$, the graph will be a horizontal line.
	
	\item If the process is a persistent fractional Brownian motion, i.e. $H> 1/2$, we will get an increasing line.
	
	\item If the process is an anti-persistent fractional Brownian motion, i.e. $H <1/2$, the result will be a decreasing line.
\end{itemize}

\bigskip


For the next graphics we used the following procedure.

 \begin{itemize}

\item The dynamic measurement of the H exponent is done with windows $ \tau $ of 250 days and lags $ \lambda$ of 5 days in complete series of indexes  of shares that quote in the BMV.

\item For the measurement of the H exponent by the $R / S$ method fragmentations of the sample of 250 daily performances were made in (r) for parts 2 (125 r), 3 (83 r), 4 (62 r), 5 (50 r), 6 (41 r), 7 (35 r), 8 (31 r), 10 (25 r), 12 (20 r) and 15 (16 r).

\item The maximum, minimum and average value of the H exponent of each sample and the coefficient $ R^2 $ were identified.

\item The date of the first measurement was recorded (corresponding to day 251 of the price series and to 250 returns) as well as the number of measurements of the Hurst exponent of each series (N).

\item We estimated the proportion of the H exponent measurements that exceeded each of the cut points marked above 0.5 and those that were below 0.5.

\end{itemize}

\begin{figure}
  \includegraphics[width=1.15\textwidth]{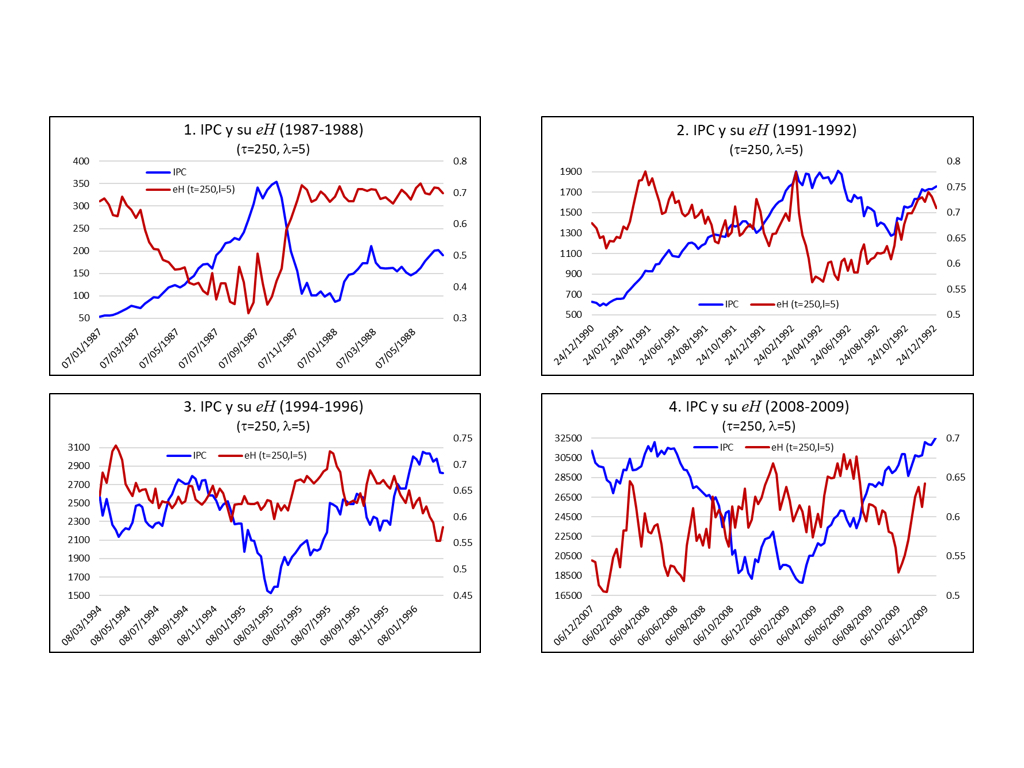}
  \caption{ IPC and the corresponding value of the H exponent for different periods with $\tau = 250$ and $\lambda = 5$. Notice the behavior of the H exponent during the respective crashes of 1987 and 2008. }
  \label{max}
\end{figure}
\section{ Comments and perspectives }

Up to this point, we do not have a (complete) equivalent of a Mean Field Theory for Financial systems. First of all, it would be necessary to identify an order parameter, like in the case of magnetic or fluid systems. It would be necessary too, to provide a well defined Gibbs' function for financial systems. 

In the present work, we have used the Hurst exponent as a parameter which gives us valuable information about the behavior of the time series of stock markets. We have studied different time series in which we used the $H$ exponent as a measure of the level of persistence in it. Our results have become encouraging. We can notice the behavior of the H exponent during deep falls. It is interesting that in the time series we have been studying, the H exponent tends to rise in the case of deep falls of asset prices. This suggests that H exponent could be used as an estimator of the level of autocorrelation and as a measure of the persistence of a long-range tendency.  

It will take some time before we have a comprehensive theory of the dynamics of financial markets. On the other hand, on the subject of phase transitions in physical systems, the book of Eugene Stanley \cite{stanley}, remains as a cornerstone. However, we still have not been able to reproduce such kind of results in the case of the stock market.  In the meantime, we have to keep working on the search for more accurate models. 

\medskip

\section{ Aknowledgments}

L.S. acknowledges support from CCC-UNAM, Mexico. 
C.S. and Z.H. acknowledge support from UAEH, Mexico.



\end{document}